\documentclass[aps,prb,twocolumn,groupedaddress,showpacs]{revtex4}
\usepackage{amsmath}
\usepackage{dcolumn}
\usepackage{amssymb}
\usepackage{amsfonts}
\usepackage{graphicx}
\usepackage[T1]{fontenc}
\usepackage{indentfirst}
\usepackage{color}

\begin{document}
\title{Efficient neural-network based variational Monte Carlo scheme for direct optimization of excited energy states in frustrated quantum systems }
\author{Tanja \DJ uri\'c}
\author{Tomislav \v{S}eva}
\affiliation{Department of Physics, Faculty of Science, University of Zagreb, Bijeni\u{c}ka c. 32, 10000 Zagreb, Croatia}

\date{\today}
\begin{abstract}
We examine applicability of the valence bond basis correlator product state ansatz, equivalent to the restricted Boltzmann machine quantum artificial neural network ansatz, and variational Monte Carlo method for direct optimization of excited energy states to study properties of strongly correlated and frustrated quantum systems. The energy eigenstates are found by stochastic minimization of the variational function for the energy eigenstates which allows direct optimization of particular energy state without knowledge of the lower energy states. This approach combined with numerous tensor network or artificial neural network ansatz wavefunctions then allows further insight into quantum phases and phase transitions in various strongly correlated models by considering properties of these systems beyond the ground state properties. Also, the method is in general applicable to any dimension and has no sign instability. An example that we consider is the square lattice $J_1$-$J_2$ antiferromagnetic Heisenberg model. The model is one of the most studied models in frustrated quantum magnetism since it is closely related to the disappearance of the antiferromagnetic order in the high-Tc superconducting materials and there is still no agreement about the properties of the system in the highly frustrated regime near $J_2/J_1=0.5$. For $J_1$-$J_2$  model we write the variational ansatz in terms of the two site correlators and in the valence bond basis and calculate lowest energy eigenstates in the highly frustrated regime near $J_2/J_1=0.5$ where the system has a paramagnetic phase. We find that our results are in good agreement with previously obtained results which confirms applicability of the method to study frustrated spin systems. 
\end{abstract} 


\maketitle
\section{Introduction}
\label{sec:Introduction}
Simulating frustrated quantum spin systems is amongst the most challenging computational tasks and is one of the central problems in condensed matter physics. 
Approximating wavefunction of the system with a tensor network (TN) or artificial neural network (ANN) ansatz and employing Monte Carlo sampling to efficiently compute expectation values proved 
recently to be a very efficient approach to study variety of strongly correlated models. \cite{Orus,Schuch,Wang3,Sfondrini,Glasser,Clark,Song,Carleo,Changlani,AlAssam,Mezzacapo1,Mezzacapo2,Neuscamman,Duric1,Duric2,Sandvik2,Jia,Shi,Deng,Chen,DasSarma} The approach can be applied to systems of any spatial dimensionality and is sign problem 
free, and therefore overcomes the limitations of two other main numerical techniques used to simulate correlated quantum spin systems, density-matrix renormalization group (DMRG) method\cite{White, Liang2, Duric3,Wang} and quantum
 Monte Carlo (QMC).\cite{Ceperley}  While the DMRG method gives very accurate results only in one dimension, the QMC suffers from the sign problem for frustrated (fermionic) quantum systems. 
 
The tensor network states (TNS) and variational Monte Carlo (VMC) approach has so far been mostly used to study the ground-state properties of various quantum correlated systems. The main reason 
for this limitation was the lack of a robust and efficient excited state variational principle analogous to the ground-state variational principle where the function that can be efficiently minimized is the energy $E(\psi)=\langle\psi|H|\psi\rangle/\langle\psi |\psi\rangle$.  Recently,
Zhao and Neuscamman proposed an efficient variational principle for the direct optimization of excited states \cite{Zhao,Neuscamman2,Blunt,Shea,Blunt2,PinedaFlores} which can be used at polynomial cost with numerous 
approximate ansatz wavefunctions. The method allows to target particular eigenstate without knowledge of the lower energy states by tuning the value of the energy shift parameter included in the variational function for the eigenstates and has so far been mostly used to study molecular excitations. In this paper we examine and confirm applicability of the mentioned variational principle combined with a suitable TNS or ANN ansatz to study properties of strongly correlated and frustrated quantum systems. 

Specifically, we calculate lowest energy eigenstates for the square lattice $J_1$-$J_2$ antiferromagnetic Heisenberg model in the highly frustrated regime near $J_2/J_1=0.5$ where the system has a paramagnetic phase. $J_1$-$J_2$  model is one of the most studied models in frustrated quantum magnetism.\cite{Wang2,Schulz1,Schulz2,Wang1,Schulz,Jiang,Sirker,Darradi,Capriotti3,Capriotti1,Li,Lou,Morita,Capriotti2} It is closely related to the disappearance of the antiferromagnetic order in high Tc superconducting materials\cite{Anderson,Lee} and is therefore of great importance. Model has also been proposed as a possible model that supports topologically ordered chiral spin-liquid state \cite{Kalmeyer, Wen1, Duric4} or $Z_2$ spin liquid state.\cite{Li,Read2,Moessner,Wen2,Yao,Hu}

The model has so far been studied using several methods, among which are, for example, exact diagonalization (ED),\cite{Schulz,Figueirido,Richter2} variational methods,\cite{Mezzacapo2, Morita,Capriotti1,Beach2,Mambrini,Chou} DMRG,\cite{Wang} the Green function Monte Carlo with stochastic reconfiguration (GFMCSR) technique\cite{Capriotti3} and the cluster update algorithm for tensor product states (TPSs).\cite{Wang2} Properties of the phase(s) in the highly frustrated regime near $J_2/J_1=0.5$ and presence of deconfined quantum critical point\cite{Sandvik2,Wang2,Merchant,Sachdev1,Senthil,Wenzel,Sachdev3} at the transition from antiferromagnetic to paramagnetic phase have been debated for decades \cite{Wang2,Schulz1,Schulz2,Wang1,Schulz,Jiang,Sirker,Darradi,Capriotti3,Capriotti1,Li,Lou,Morita,Kalmeyer, Wen1, Duric4,Read2,Moessner,Wen2,Yao,Hu} and there is still no general agreement.  

Here we demonstrate that VMC method for direct optimization of excited energy states combined with appropriate TN or ANN ansatz can provide further insight into quantum phases and phase transitions in complex models such as frustrated Heisenberg $J_1$-$J_2$ model. The method is applicable in any dimension, allows studying properties of the system beyond the ground state properties and in general has no sign instability. 

Our calculations are performed with correlator product state (CPS) ansatz\cite{Changlani,AlAssam,Duric1,Duric2} in the valence bond (VB) basis\cite{Morita,Capriotti1,Li, Sutherland, Beach, Liang,Sandvik1,Lou,Read,Anderson,Tang,Beach2} as an ansatz for the energy eigenstates. Here the CPS ansatz is built from two-site correlators associated with the pairs of sites. The ansatz is equivalent to the restricted Boltzmann machine (RBM) representation of the wavefunction where the number of hidden units equals to the number of different pairs of sites. RBMs are types of generative stochastic artificial neural networks (ANNs)\cite{Glasser,Clark,Carleo,Jia,Shi,Deng,Chen,DasSarma} that can learn a distribution over the set of their inputs. The inputs here are spin configurations and the wavefunction corresponds to the complex probability distribution that the network tries to approximate. Correlations in ANNs are included by hidden units and are nonlocal in space. Due to their non-local geometry  ANNs can describe some of the states that can not be described by traditional TNSs, for example, chiral spin liquid states or lattice fractional quantum Hall states. 

Variational function for the energy eigenstates is minimized using stochastic optimization scheme \cite{AlAssam,Duric1,Duric2,Sandvik2,Lou,Robbins,Spall} which requires knowledge only of the first derivatives of the variational function with respect to the variational parameters in the CPS ansatz. We calculate lowest energy eigenstates in total spin zero sector for the system sizes with $N=36$ and $64$ lattice sites, and with periodic boundary conditions. Our results for the energy gap between the first excited energy state and the ground state in total spin zero sector is in good agreement with results obtained previously with other methods. This demonstrates general applicability of the method to study properties of complex interacting many-body systems.

We also note that statistical error present in the stochastic algorithm can result in significant error for energy eigenstates for smaller system sizes. However, influence of the error can be controlled by increasing the system size and does not affect results for the energy gap where the error cancels when subtracting values of the eigenenergies. 

The paper is organized as follows. In Sec. \ref{sec:RVBCPS} we define CPS variational ansatz in terms of two-site correlators and in the VB basis. In Sec. \ref{sec:StochasticMinimization} we describe stochastic optimization scheme for efficient minimization of the variational function for energy eigenstates. Our numerical results for the system sizes $N=L\times L$ lattice sites with $L=6$ and $8$ and periodic boundary conditions are presented in Sec. \ref{sec:NumericalResultsJ1J2}. In the final section Sec.\ref{sec:Conculsions} we draw our conclusions, summarize results and discuss possible directions for future research. 

\section{Valence bond basis correlator product states}
\label{sec:RVBCPS}

We consider the square lattice $J_1$-$J_2$ antiferromagnetic Heisenberg model
\begin{equation}\label{eq:H}
H = J_1\sum_{\langle i,j \rangle}\vec{S}_i\vec{S}_j + J_2\sum_{\langle\langle i,j\rangle\rangle}\vec{S}_i\vec{S}_j,
\end{equation}
where $\vec{S}_i$ are spin-$1/2$ operators, and $J_1$ and $J_2$ are antiferromagnetic couplings for neighboring and next-neighboring sites, respectively.

In general, in the usual basis of the $S_z^{tot}=\sum_{i=1}^NS_i^z$ eigenstates, an eigenstate of the Hamiltonian (\ref{eq:H}) can be written in the form
\begin{equation}\label{eq:wf1}
|\psi\rangle = \sum_{\{\vec{s}\}} W(\{\vec{s}\})|\{\vec{s}\}\rangle,
\end{equation}
where $|\{\vec{s}\}\rangle=|s_1,...,s_N\rangle$ and $s_i\in\{-1/2,1/2\}$ are eigenvalues of the local $S_i^z$ operator. Since the total magnetization along the $z$ axis is a good quantum number, i. e. the Hamiltonian (\ref{eq:H}) commutes with the operator $M_z=S_z^{tot}$, 
\begin{equation}\label{eq:HMz}
\left[H, M_z\right]=0,
\end{equation}
all calculations can be performed in the canonical ensemble, i.e. for a chosen $M_z$ sector. The eigenstates can then also be written in the form 
\begin{equation}\label{eq:wf2}
|\psi\rangle  _{M_z} = \sum_{\{\vec{s}\}} W(\{\vec{s}\})P_{M_z}|\{\vec{s}\}\rangle, 
\end{equation}
where the projection operator $P_{M_z}$ projects to the spin configurations with $\sum_{i=1}^NS_i^z=M_z$. 

 The ground state and the lowest excited energy state in the highly frustrated regime near $J_2/J_1 =0.5$ are in the $M_z=0$ sector \cite{Morita,Schulz1,Schulz2} and we therefore restrict our calculations to $M_z=0$. Then the most suited basis for the spin-rotationally invariant Hamiltonians is the VB basis.\cite{Morita,Capriotti1,Li, Sutherland, Beach, Liang,Sandvik1,Lou,Read,Anderson,Tang} A VB configuration $|\alpha\rangle$ 
\begin{equation}\label{eq:VBconfiguration}
|\alpha\rangle = |(i_1^\alpha,j_1^\alpha)(i_2^\alpha,j_2^\alpha)...(i_{N/2}^\alpha,j_{N/2}^\alpha)\rangle,
\end{equation}
is a product of two-spin singlets 
\begin{equation}\label{eq:TwoSpinSinglet}
(i,j)=\frac{1}{\sqrt{2}}\left(|\uparrow_i\downarrow_j\rangle-|\downarrow_i\uparrow_j\rangle\right),
\end{equation}
and any total singlet state with $S_z^{tot}=0$ can be written in the VB basis 
\begin{equation}\label{eq:VBbasis}
|\psi\rangle =\sum_{\alpha} W({\alpha})|\alpha\rangle,
\end{equation}
where VB configurations $\alpha$ correspond to all possible pairings of $N$ spins into $N/2$ valence bonds. 

\begin{figure}[t!]
\includegraphics[width=\columnwidth]{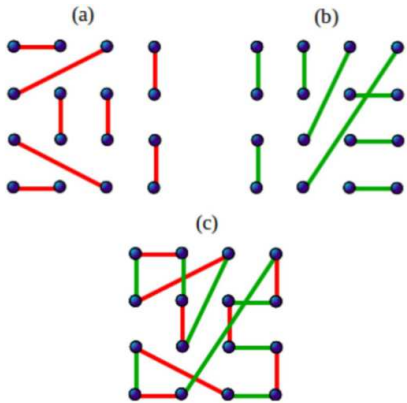}
\caption{\label{Fig:TranspositionGraph}Transposition graph (c) $\langle\alpha|\beta\rangle$ of two valence bond states (a) $|\alpha\rangle$ and (b) $\langle\beta|$. The valence bond basis (\ref{eq:VBconfiguration}) is an overcomplete basis and the overlap between two valence bond states is $\langle\alpha|\beta\rangle = \pm 2^{N_l-N/2}$ where $N_l$ is the number of loops in the transposition graph and $N$ is the number of lattice sites. Here $N=16$ and $N_l=2$. 
}
\end{figure}
The VB basis is an overcomplete basis and the overlap between the VB states is \cite{Capriotti1,Li,Liang,Sandvik1,Lou,Read,Anderson}
\begin{equation}\label{eq:VBoverlap}
\langle\alpha|\beta\rangle = (-1)^{n_{\alpha}+n_{\beta}}2^{N_l-N/2}=\pm 2^{N_l-N/2},
\end{equation}
where $N_l$ is the number of loops in the transposition graph obtained when the VB states $|\alpha\rangle$ and $|\beta\rangle$ are superimposed (FIG. \ref{Fig:TranspositionGraph}) and the overall sign depends on the convention for assigning directions to the bonds ($n_{\alpha}$ and $n_{\beta}$ denote the number of valence bonds in $|\alpha\rangle$ and $|\beta\rangle$ in the opposite direction than defined to be + direction).  Because of the overcompleteness of the VB basis the expansion coefficients $W(\alpha)$ in Eq. (\ref{eq:VBbasis}) are not unique. This is however not a problem for any practical calculations. 

Also, the trial wavefunction doesn't have to be constructed from the largest possible set of VB states in which all spins are joined by valence bonds in all possible ways. A more restricted (and still overcomplete) basis can be obtained by dividing the system into two groups of sites, A and B, and keeping only VB states with bipartite bonds which connect sites from different groups A and B. The overcompleteness property can be written as $(i,k)(j,l)\rightarrow (i,j)(k,l)-(i,l)(k,j)$\cite{Beach} and VB states with non-bipartite AA and BB bonds can be therefore written in terms of VB states with AB bonds. 

For a bipartite lattice typical choice of the sites A and B corresponds to the two sublattices in the bipartite lattice. For a square lattice AB labels are usually assigned to form checkerboard or collinear patterns. If in addition the direction of each singlet $(i,j)$ in a VB state is fixed such that $i \in A$ and $j\in B$ it can be shown that all expansion coefficients $W(\alpha)$ can be taken to be real and positive.\cite{Sutherland,Liang} That corresponds to Marshall's sign rule\cite{Marshall,Schollwock} in the absence of frustration when the wavefunction is written in the standard basis of eigenstates of the $S_i^z$ operator (\ref{eq:wf1}).

The Marshall's sign rule exists in two limits $J_2/J_1=0$ and $J_2/J_1=\infty$ and it can be shown that the sign rule survives the frustration in the $J_1$-$J_2$ model on the square lattice for a relatively large range of the parameter $J_2/J_1$ values away from the points  $J_2/J_1=0$ and $J_2/J_1=\infty$.\cite{Richter,Voigt} Also, for any $J_2/J_1$ exists in principle a positive-definite expansion of the wavefunction in the VB basis, since $W(\alpha)$ can be made positive by simply reversing the order of the indices in one singlet in that particular state. However, in general there is no practically useful rule for fixing the order. 

Within the CPS approach \cite{AlAssam,Duric1,Duric2} the coefficients in Eq. (\ref{eq:wf1}) or Eq. (\ref{eq:VBbasis}) are written in terms of correlator coefficients associated with groups of sites. The CPS ansatz can then be used as a basis for VMC simulations where the coefficients are optimized using one of the efficient optimization methods, for example stochastic optimization scheme described in the following section, which requests only
the first energy derivatives.

Here we consider two-site CPS where a correlator is associated with a pair of sites 
\begin{equation}\label{eq:wf_CPS}
W(\alpha)=(-1)^{n_{\alpha}}\prod_{i,j}C_{ij}.
\end{equation} 
In the previous calculations for the $J_1$-$J_2$ model computational cost was reduced by assuming symmetries for the coefficients in an ansatz wavefunction, for example $2 \times 2$ sublattice structure and translational invariance of the coefficients in terms of the sublattice period.\cite{Morita} Since it is better to use a flexible ansatz wavefunction without any constraint on coefficients $C_{ij}$ we do not take into consideration any symmetries and we also do not impose any constraint between $C_{ij}$ and $C_{ji}$. Amplitudes $C_{ij}$ can than be taken to be real and positive. We consider $L\times L$ square lattices with periodic boundary conditions where $i=(x,y)$ and $x,y=1,... ,L$. For the $L\times L$ square lattice, with $N =L^2$ lattice sites, the number of coefficients $W_{\alpha}$ is $N!$, and the two-site CPS ansatz (\ref{eq:wf_CPS}) leads to $N(N-1)$ variational parameters. 

CPS states in VB basis can be used to describe both ordered and disordered phases.\cite{Liang, Sandvik1} Neel ordered ground state with spin correlations decaying with a power law as a function of distance requires long-range amplitudes $C_{ij}$ while a disordered  state requires larger decay rate of amplitudes $C_{ij}$ with distance $|i-j|$.

Also this CPS ansatz is equivalent to the RBM representation of the wavefunction with $ M = N(N-1)/2$ hidden units.\cite{Glasser,DasSarma} Boltzmann machines are types of generative stochastic artificial neural networks that can learn a distribution over the set of their inputs.\cite{Glasser,Clark,Carleo,Jia,Shi,Deng,Chen,DasSarma} Here the network inputs are spin configurations and the wave-function corresponds to a complex probability distribution that the network tries to approximate.

\section{Stochastic optimization of the variational function for energy eigenstates}
\label{sec:StochasticMinimization}

So far variational calculations have been mostly restricted to studying the ground state properties of various systems with the energy as a function that can be efficiently minimized,
\begin{equation}\label{eq:En}
 E(\psi)=\frac{\langle\psi|H|\psi\rangle}{\langle\psi|\psi\rangle},
\end{equation}
where $H$ is the Hamiltonian of the system. Optimization of an excited energy state would then require knowledge of all energy states with energy lower than the energy of the chosen excited state.

However, Zhao and Neuscamman \cite{Neuscamman2,Blunt,Shea,Blunt2,PinedaFlores} recently introduced an efficient variational principle for direct optimization of excited states that does not require knowledge of lower energy states. They have defined a function 
\begin{equation}\label{eq:Omega}
\Omega({\psi,\omega})=\frac{\langle\psi|(\omega-H)|\psi\rangle}{\langle\psi|(\omega-H)^2|\psi\rangle}=\frac{\omega-E}{(\omega-E)^2+\sigma^2},
\end{equation}
where
\begin{equation}\label{eq:sigma}
\sigma^2=\frac{\langle\psi|(H-E)^2|\psi\rangle}{\langle\psi|\psi\rangle}
\end{equation}
is the variance, whose global minimum is an excited energy state with the energy immediately above the energy shift $\omega$ that is placed between distinct eigenvalues of $H$.

To efficiently evaluate and optimize $\Omega(\psi,\omega)$ using variational Monte Carlo scheme, $\Omega(\psi,\omega)$ is rewritten in the form
\begin{equation}\label{eq:Omega2}
\Omega(\psi, \omega)=\frac{\sum_{m}P(m)w_m}{\sum_{m}P(m)w_m^2},
\end{equation}
where $m=(\alpha,\alpha')$ denotes a pair of two VB configurations $|\alpha\rangle$ and $|\alpha'\rangle$,  
\begin{equation}\label{eq:wm}
w_m\equiv\frac{\langle \alpha'|(\omega - H)|\alpha\rangle}{\langle \alpha'|\alpha\rangle},
\end{equation}
and 
\begin{equation}\label{eq:Pm}
P(m)=\frac{W_{\alpha'}W_{\alpha}\langle\alpha'|\alpha\rangle}{\sum_{\alpha,\alpha'}W_{\alpha'}W_{\alpha}\langle\alpha'|\alpha\rangle}.
\end{equation}
$P(m)$ is always positive since the coefficients $C_{ij}$ are taken here to be real and positive. 
\begin{figure}[t!]
\includegraphics[width=\columnwidth]{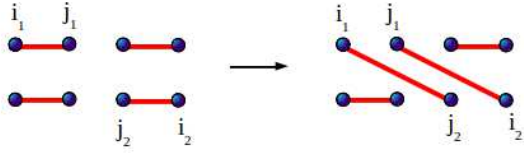}
\caption{\label{Fig:TwoBondUpdates}Local two-bond updates. First site $i_1$ is randomly chosen, then one of four diagonal neighbor
sites $i_2$ is chosen randomly and the ends of bonds are exchanged: $(i_1 , j_1 )(i_2 , j_2 ) \rightarrow (i_1 , j_2 )(i_2 , j_1 )$ . The reconfiguration is accepted with probability (\ref{eq:Pacc1}).  
}
\end{figure}

Similarly to the overlap of two VB configurations (\ref{eq:VBoverlap}) matrix elements of relevant operators can be typically calculated by considering loops in the transposition graph.\cite{Sutherland,Liang,Sandvik1,Beach}The spin-spin correlations and $\langle\alpha'|H|\alpha\rangle/\langle\alpha'|\alpha\rangle$ can be computed from the relation
\begin{equation}\label{eq:MatrixElements}
\frac{\langle\alpha'|\vec{S}_i\vec{S}_j|\alpha\rangle}{\langle\alpha'|\alpha\rangle}=\frac{3}{4}\phi_i\phi_j\delta_{\lambda_i,\lambda_j},
\end{equation}
where $\phi_i=+1$ if sites $i$ and $j$ belong to two different groups of sites (A and B), $\phi_i=-1$ if sites $i$ and $j$ belong to the same group of sites (A or B), and $\delta_{\lambda_i,\lambda_j}=1$ if $\lambda_i=\lambda_j$ and zero otherwise. Here $\lambda_i$ is a label for the loop to which site $i$ belongs, and matrix element (\ref{eq:MatrixElements}) vanishes if sites $i$ and $j$ belong to different loops.  

Within VMC scheme the phase space considered as ensemble of pairs $m=(\alpha,\alpha')$ is summed over according to probability distribution $P(m)$ and 
\begin{equation}\label{eq:OmegaVMC}
\Omega_{VMC}(\psi,\omega)=\frac{\sum_{m\in\zeta}w_m}{\sum_{m\in\zeta}w_m^2},
\end{equation}
where elements of $\zeta$ are sampled via a Metropolis walk. In evaluating the sums in equation (\ref{eq:OmegaVMC}) a new pair of valence bonds $m'$ is generated starting from a valence bond pair $m$ with a chosen update scheme and Metropolis acceptance probability for such update is 
\begin{equation}\label{eq:Pacc}
P_{acc}=\min\left[\frac{W_{m'}}{W_m}2^{\Delta N_l},1\right]
\end{equation}
where $W_m=W_{\alpha'}W_{\alpha}$ and $\Delta N_l=N_{l}(m')-N_l(m)$ denotes change in the number of loops in the transposition graph (FIG. \ref{Fig:TranspositionGraph}).
\begin{figure}[b!]
\includegraphics[width=\columnwidth]{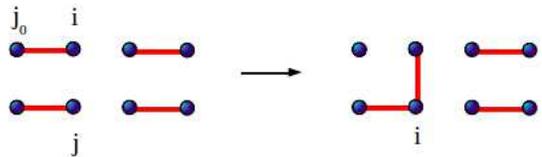}
\caption{\label{Fig:BondLoopUpdates}Non-local bond-loop updates. The first step in a bond-loop update discussed in the text which creates two defects (one site with two bonds and one empty site). Since such defects can not be present in a valence bond state one of these defects is further moved by subsequent bond moves until it annihilates with the second defect and the loop closes. 
}
\end{figure}
Here Monte Carlo sampling is performed by non-local bond-loop updates \cite{Sandvik1} instead with local two-bond updates \cite{Liang, Sandvik1} since non-local updates sample the phase space in the highly frustrated regime much more efficiently. In local two-bond update a new configuration in the MC move is generated by exchanging the ends of two bonds $(i_1,j_1)$ and $(i_2,j_2)$ as illustrated in FIG. \ref{Fig:TwoBondUpdates}. First site $i_1$ is randomly chosen, then one of four diagonal neighbor sites $i_2$ is chosen randomly and the ends of bonds are exchanged $(i_1,j_1)(i_2,j_2)\rightarrow(i_1,j_2)(i_2,j_1)$. The Metropolis acceptance probability is then 
\begin{equation}\label{eq:Pacc1}
P_{acc}=\min\left[\frac{C_{i_1,j_1}C_{i_2,j_2}}{C_{i_1,j_2}C_{i_2,j_1}}2^{\Delta N_l},1\right]
\end{equation}
and $\Delta N_l$ can be +1,-1 or 0 corresponding to the cases when two loops join in one, one loop splits into two, or number of loops is preserved. 
However, local two-bond updates usually involve long bonds through reconfiguration and the sampling process becomes inefficient when long bond amplitudes have small weights. Since the values of longer bond amplitudes decrease with decrease of antiferromagnetic ordering non-local bond updates can be much more efficient sampling scheme, particularly in the highly frustrated regime where disappearance of antiferromagnetic ordering is predicted. Within bond-loop update scheme one end of a randomly chosen bond is moved resulting in two defects that correspond to an empty site and one site with two bonds (FIG. \ref{Fig:BondLoopUpdates}). Since such defects can not be present in a VB state one of these defects is further moved by subsequent bond moves until it annihilates with the second defect and the loop closes. 

In the algorithm the bonds are represented by an array of links between sites, here denoted by $v$, such that if sites $i$ and $j$ are connected by a VB then $v(i)=j$ and $v(j)=i$. In the bond-loop update algorithm, starting from a randomly selected lattice site $j_0$ with $v(j_0)=i$, a new lattice site $j$ is chosen according to probability distribution 
\begin{equation}\label{eq:BondLoopUpdates}
P_{ij}=\frac{C_{ij}}{\sum_jC_{ij}},
\end{equation}
proportional to corresponding weight $C_{ij}$. After the bond emerging from $i$ is moved to an acceptable $j$, $v(i)$ changes from $v(i)=j_0$ to $v(i)=j$, and the original link between sites $i$ and $j_0$ is destroyed and no longer needed.

The original site $j_0$ now has no bond attached to it (unless $j=j_0$ which immediately terminates bond-loop update and a new loop update starts from a different randomly chosen site) and the new site $j$ has two VBs on it, corresponding to two defects in VB configuration. To remove such defects which can not be present in a VB state the end of the old bond $i$ is moved by repeating the same steps as for the initial bond move only with $j_0$ replaced by $j$. This procedures are repeated until it happens that $j=j_0$ which results in annihilation of the double-bond and no-bond defects and closing of the loop. Since the loops can be large bond-loop updates can be much more efficient than local two-bond updates. 

To optimize $\Omega(\psi,\omega)$ we further use a stochastic optimization scheme \cite{AlAssam,Duric1,Duric2,Sandvik2,Lou,Robbins,Spall} which requests only knowledge of the signs of the first derivatives of $\Omega(\psi,\omega)$ with respect to parameters $C_{ij}$ which are updated according to 
\begin{equation}\label{eq:update_Cij}
\ln C_{ij} \rightarrow \ln C_{ij} -r\delta(k)\mathrm{sign} \left(\frac{\partial\Omega}{\partial C_{ij}}\right),
\end{equation}
where $r\in [0,1)$ is a random number and $\delta(k)$ is the optimization step for given iteration $k$. Without random number this kind of update scheme is known as Manhattan learning \cite{Peterson,Leen} previously introduced in the context of neural networks. Here random number is introduced because it was shown that it speeds up the convergence.\cite{Lou}

Similarly as in simulated annealing methods,\cite{Kirkpatrick} the optimization step $\delta(k)$ is reduced in each iteration $k$ to reach the optimum solution. Here the annealing scheme that ensures convergence of the method is 
\begin{equation}\label{eq:delta_k}
\delta(k)=\delta_0\cdot\frac{1}{k^\nu}
\end{equation}
with $0.5 < \nu< 1$, as demonstrated previously for the energy minimization.\cite{Lou} We find that taking $\delta_0=0.5$ and $\nu=0.75$ works well. Alternatively, a geometric form $\delta(k)=\delta_0\nu^k$ for the annealing scheme can also be used with $\nu=1-\epsilon$ and $\epsilon\ll1$.\cite{Lou}

First derivatives $\partial \Omega/\partial C_{ij}$ can be efficiently evaluated using variational Monte Carlo scheme using the following expression 
\begin{equation}\label{eq:FirstDerivatives}
\frac{\partial{\Omega}}{\partial C_{ij}} = \frac{1}{\langle\mathcal{O}_2\rangle}\frac{\partial\langle\mathcal{O}_1\rangle}{\partial C_{ij}}-\frac{\langle\mathcal{O}_1\rangle}{\langle\mathcal{O}_2\rangle^2}\frac{\partial\langle{O}_2\rangle}{\partial{C_{ij}}},
\end{equation}
where $\langle\mathcal{O}_1\rangle=\langle\psi|\omega -H |\psi\rangle/\langle\psi|\psi\rangle\equiv\langle\omega-H\rangle$ and $\langle\mathcal{O}_2\rangle=\langle(\omega-H)^2\rangle$, 
\begin{equation}\label{eq:FirstDerivatives2}
\frac{\partial\langle\mathcal{O}_k\rangle}{\partial C_{ij}}=\langle\Delta_{ij}\mathcal{O}_k\rangle-\langle\Delta_{ij}\rangle\langle\mathcal{O}_k\rangle
\end{equation}
for $k=1$ or $2$, and 
\begin{equation}\label{eq:Delta}
\Delta_{ij}=\frac{1}{C_{ij}}\frac{\partial W_m}{\partial C_{ij}}=\frac{b_{ij}}{C_{ij}}. 
\end{equation}
Here $b_{ij}$ denotes the number of times the coefficient $C_{ij}$ appears in the product $W_m=W_{\alpha}W_{\alpha'}$ where the amplitude $W_{\alpha}$ for the VB configuration $|\alpha\rangle$ is given by equation (\ref{eq:wf_CPS}). Since we do not include any symmetries a coefficient $C_{ij}$ appears in each VB configuration only once. Therefore $b_{ij}=1$ if $C_{ij}$ appears in only one of the VB configurations ($\alpha$ or $\alpha'$) or $b_{ij}=2$ if $C_{ij}$ appears in both VB configurations ($\alpha$ and $\alpha'$). The first derivatives can be calculated from the same sample as $\Omega_{VMC}(\psi,\omega)$ (\ref{eq:OmegaVMC}) obtained by a Markov chain in the Metropolis algorithm. 

The variational algorithm starts from randomly chosen values for the coefficients $C_{ij}$ (between $0$ and $1$), then $\Omega$ and its gradient vector is evaluated for particular value of the parameter $\omega$ and all coefficients $C_{ij}$ are updated according to (\ref{eq:update_Cij}). In each iteration the same procedure is repeated, starting from the coefficients $C_{ij}$ from the previous iteration, until convergence of $\Omega$ is reached. The value of energy $E$ is calculated using coefficients obtained by minimizing $\Omega$. 

In each iteration $k$ the variational function $\Omega$ and its derivative are estimated from $F(k)\times N$ sampled values where $N$ is the number of lattice sites. $F(k)$ is called the number of sweeps per sample. In each sweep a random lattice site is chosen and a move to a new configuration obtained by a bond-loop update is proposed $N$ times. In addition to careful tuning of the gradient step $\delta(k)$ to achieve the convergence, the number of sweeps $F(k)$ per iteration is increased to reduce effects of noise on the calculation of the first derivatives. Namely, the derivatives become smaller as the $\Omega$ minimum is approached and require more sampled values in order not to be dominated by noise. Here, the number of sweeps is increased linearly for each iteration, $F(k)=F_0\times k$. 

Also, the procedure of evaluating $\Omega$ and updating the coefficients $C_{ij}$ (\ref{eq:update_Cij}) is repeated $G (k)= G_0\times k$ times where increasing $G$ corresponds to a slower cooling rate. Here we take $G_0=20$, $F_0=20$ for $L=6$ and $F_0=10$ for $L=8$. The minimization routine is performed for $100$ iterations and after the minimization is complete the values of $\Omega$, and corresponding energy $E$ 
for each value of $\omega$, are calculated by repeating the procedure for a single iteration with zero step size and large F and G to obtain more accurate estimates of $\Omega$ and $E$. 

\section{Numerical results for the square lattice $J_1$-$J_2$ antiferromagnetic Heisenberg model } 
\label{sec:NumericalResultsJ1J2}
\begin{figure}[t!]
\includegraphics[width=\columnwidth]{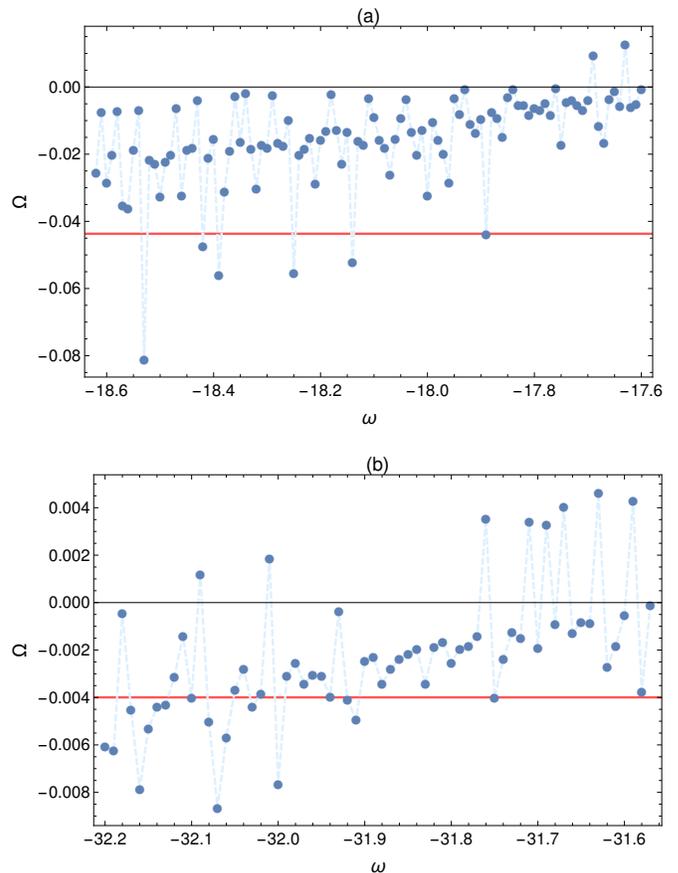}
\caption{\label{Fig:Omega} $\Omega$ as a function of the energy shift parameter $\omega$ for the ground state and the first excited energy state ($M_z=\sum_{i=1}^{N} S_i^z=0$ sector) for $J_2/J_1=0.5$ ($J_1$ is set to $J_1=1$) and the system sizes $N=L\times L$ with (a) $L=6$, (b) $L=8$ and with periodic boundary conditions. Here  $|\psi^{ansatz}\rangle$ is optimized using stochastic optimization method to minimize $\Omega(\psi^{ansatz},\omega)$ for each value of $\omega$. Red lines denote minimal value of $\Omega(\omega)$, $\Omega_1^{min}$, for $\omega$ within the range $E_0\leq\omega\leq E_1$. Energy of the first excited energy state obtained by optimization procedure is calculated using the optimized ansatz that minimized $\Omega$ at $\omega_1^{min}$ where $\Omega(\omega_1^{min})=\Omega_1^{min}$.  
}
\end{figure}

\begin{figure}[t!]
\includegraphics[width=\columnwidth]{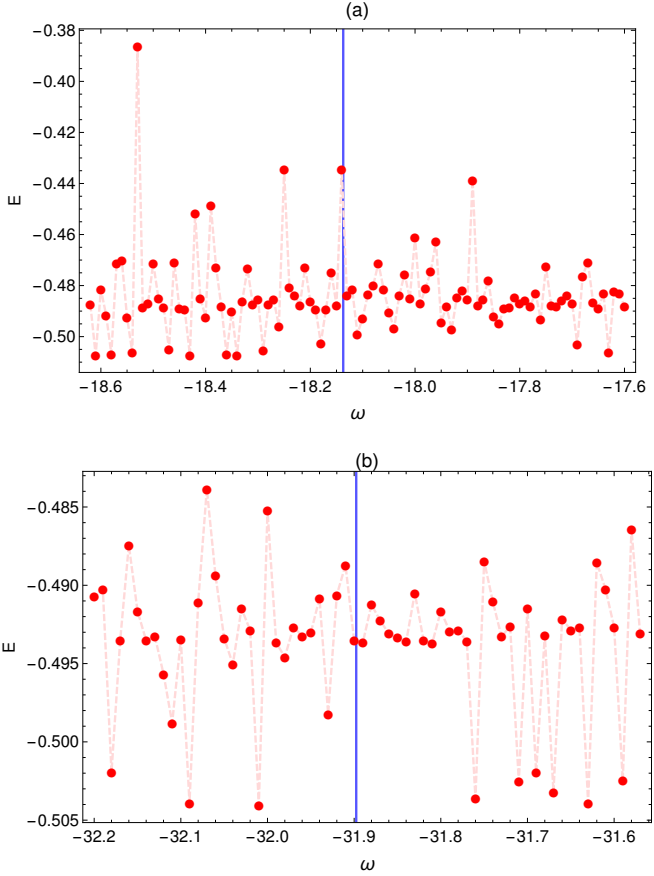}
\caption{\label{Fig:Energy}Energy $E$ as a function of the energy shift parameter $\omega$ calculated by taking the optimized ansatz that minimizes $\Omega(\psi^{ansatz},\omega)$ for each value of $\omega$. The results are for the system sizes $N=L\times L$ with (a) $L=6$, (b) $L=8$ and periodic boundary conditions. Here $J_2/J_1=0.5$ and $J_1$ is set to $J_1=1$. Blue lines denote the ground state energies obtained by exact diagonalization and the cluster update algorithm for tensor product states.\cite{Wang2}  
}
\end{figure}
\begin{figure}[b!]
\includegraphics[width=\columnwidth]{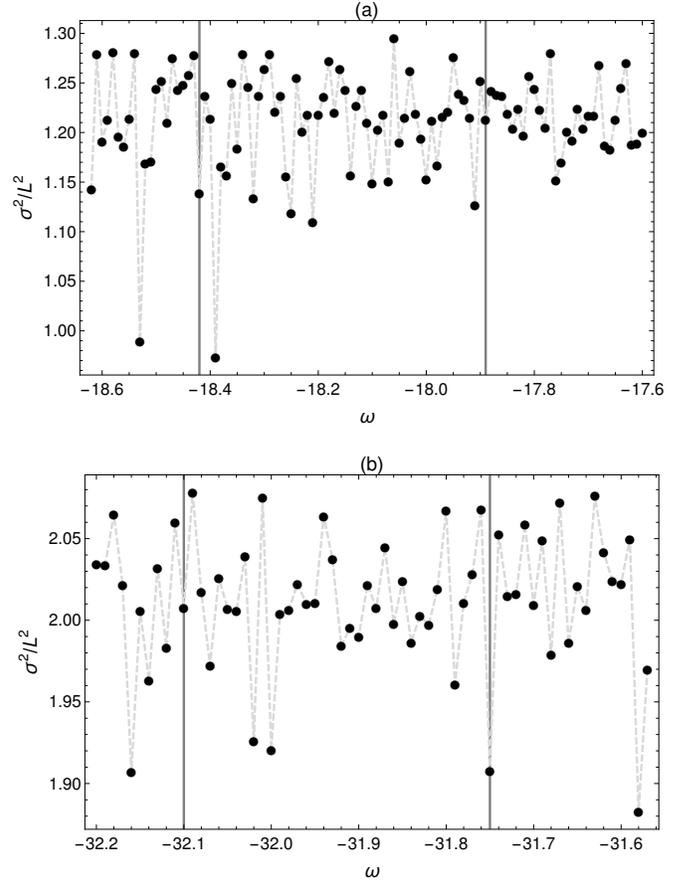}
\caption{\label{Fig:Sigma}Variance $\sigma^2$ as a function of the energy shift parameter $\omega$ for $J_2/J_1=0.5$ ($J_1=1$) and the system sizes $N=L\times L$ with (a) $L=6$, (b) $L =8$ and periodic boundary conditions. Gray lines denote values of $\omega$ for which the energy eigenvalues for the ground and first excited states are calculated. 
}
\end{figure}
Before proceeding to numerical results for the square lattice $J_1$-$J_2$ antiferromegnatic Heisenberg model it is important to clarify consequences of using approximate ansatz for the wavefunctions that correspond to the energy eigenstates. For the exact wavefunction $\sigma_n=0$ and $\Omega(\psi,\omega)$ diverges at $\omega_n=E_n$. However, for an ansatz wavefunction variance $\sigma$ has a nonzero value and the optimized ansatz, its energy and value of $\Omega(\psi,\omega)$ depend on particular choice of $\omega$. 

If the approximate ansatz wavefunction is very close to the exact wavefunction energy dependence on precise choice of $\omega$ is small and the function $\Omega(\psi,\omega)$ has a finite minimum near $\omega_n=E_n-\sigma_n$ for the n-th eigenstate with energy $E_n$.\cite{Zhao} Value of $\omega_n=E_n-\sigma_n$ corresponds to the analytic solution for the minimum of $\Omega(\psi,\omega)$ when $E$ and $\sigma$ are held fixed. Consequently for a finite and small $\sigma_n$ the value of $\omega_n$ at which minimum of $\Omega(\psi,\omega)$ changes states no longer occurs at the energy of lower energy eigenstate and is shifted downward to the value of $\omega$ close to $\omega_n=E_n-\sigma_n$. However, in the cases where the wavefunction approximation leads to a larger value of $\sigma$ and the ansatz is not very close to the exact eigenstate wavefunction a strong dependence of the energy on $\omega$ may arise. Therefore, as proposed by Zhao and Neuscamman\cite{Zhao} $\omega$ should be chosen to minimize $\Omega$ for particular eigenstate $E_n$.

It is also important to point out that optimizing the energy $E=\langle\psi|H|\psi\rangle/\langle\psi|\psi\rangle$ and $\Omega(\psi,\omega)$ is different since quality of the wavefunction depends both on energy and its variance. Therefore even for the ground state, which can also be obtained by minimizing energy $E$, we need to take the value $E_0$ obtained by minimizing $\Omega$ to calculate the energy difference $E_n-E_0$. The value of $E_0$ obtained by minimizing $E$ corresponds to the minimum of $\Omega(\psi,\omega\rightarrow\-\infty)$.

Our results obtained by minimizing $\Omega$ for $J_2/J_1=0.5$ (where $J_1$ is set to $J_1=1$) and the system sizes $N=L\times L$ with $L=6$, $8$ and periodic boundary conditions are shown in FIG. \ref{Fig:Omega} - FIG. \ref{Fig:Sigma}. Statistical error present in the stochastic algorithm is controlled by increasing the system size \cite{AlAssam, Duric2} since having a larger number of parameters allows the optimization method more freedom in finding the minimum of $\Omega(\psi,\omega)$ for a given value of $\omega$ and consequently better estimates for the energy eigenvalues. It is therefore difficult to obtain good estimates of $\Omega$ and $E$ for small system sizes. This is clearly visible in our results since we obtain much better energy estimates for $L=8$ than for $L=6$ as it will be clarified further in this section. However, we obtain quite good estimates for energy gaps in both cases since the statistical error equivalently affects calculation of all energy states and cancels in the energy gap calculation. 

We calculate the energy of the first excited energy state as energy of the optimized ansatz at the value of $\omega$ within $E_0\leq\omega\leq E_1$ range where the optimized value of $\Omega(\psi,\omega)$ is minimal. For $L=6$ the ground state energy obtained by exact diagonalization is $E_0(L=6)/L^2= - 0.50380965$\cite{Wang2,Morita} and we find that for $\omega\geq E_0$ value of $\Omega(\psi,\omega)$ is minimal at $\omega_1 = - 17.89$. Energy of the corresponding optimized ansatz at $\omega_1 =-17.89$ is $E_1^{ansatz}/L^2= -0.4386$. The value of $E_0^{ansatz}$ that corresponds to $\Omega$ minimization is determined by minimizing $\Omega(\psi,\omega)$ for a range of values $\omega < E_0$ and then finding the value $\omega_0$ such that $\Omega(\psi_0,\omega_0)\approx \Omega(\psi_1,\omega_1)$. For $L=6$ we find that $\omega_0=-18.42$ and $E_0^{ansatz}/L^2=-0.45168$. The ground state energy obtained by $\Omega$ minimization therefore significantly differs from the exact ground state energy for the smaller system size with $L=6$ with $(E_0^{ED}-E_0^{ansatz})/E_0^{ED} \approx 0.1$. However, the energy gap $\Delta = E_1^{ansatz}-E_0^{ansatz}\approx 0.471$ is very close to the value obtained by exact diagonalization and GFMCSR technique.\cite{Capriotti3, Morita}

For $L=8$ influence of the statistical error in the stochastic algorithm is much smaller and therefore better estimates for $E_0$ and $E_1$ are obtained from minimization of $\Omega$. For the first excited energy state minimum of $\Omega$ is found at $\omega=-31.75$ and corresponding energy estimate is $E_1^{ansatz}/L^2=-0.488457$. The value of $\omega_0$ for which $\Omega(\psi_0,\omega_0)\approx \Omega(\psi_1,\omega_1)$ is $\omega_0 = -32.1$ for $L=8$ and the ground state energy estimate is $E_0^{ansatz}/L^2=-0.493472$. This value differs only by $\approx 1\%$ from the value obtained by the cluster update algorithm for TPSs\cite{Wang2} ($E_0^{TPS}/L^2= -0.4984(2)$, $(E_0^{ansatz}-E_0^{TPS})/E_0^{TPS}\approx0.00993$). For the energy gap we obtain $\Delta = E_1^{ansatz}-E_0^{ansatz}\approx 0.321$ ($\Delta/L^2 \approx 0.005$) which is in agreement with previously obtained results calculated with GFMCSR technique\cite{Capriotti3} and VMC combined with the quantum number projection.\cite{Morita}

We also note that variance (FIG.\ref{Fig:Sigma}) of the optimized ansatz wavefunction obtained by minimization of $\Omega$ is quite large. This could perhaps be corrected by choosing a more complex ansatz wavefunction that would better mimic correlations built into exact eigenvalues, which is one of the directions for future research.   

\section{Conclusions}
\label{sec:Conculsions}
We have investigated applicability of the VMC method for direct optimization of energy eigenstates combined with appropriate ansatz for many-body wavefunction to study properties of complex strongly correlated and frustrated quantum systems. To demonstrate the applicability of the model we have calculated energy gaps for the square lattice $J_1$-$J_2$ frustrated Heisenberg model related to high-$T_c$ superconducting materials. Our results are in good agreement with results obtained previously by other methods, particularly for larger system size where the influence of the statistical error included in the stochastic optimization scheme used in our calculations is smaller. The results therefore confirm applicability of the method to study properties of strongly correlated systems beyond the ground state properties. This allows further insight into quantum phases and phase transitions in various correlated models. 

Directions of our future research are more detailed calculations of the system properties for several values of the parameter $J_2/J_1$ close to quantum critical point which is argued to be a deconfined quantum critical point.\cite{Wang2}We also plan to perform further calculations with several different, more complex ansatz states that could approximate correlations built into exact eigenstates better than the ansatz used in the present calculation (for example RBM wavefunction with increased number of hidden units).

\begin{acknowledgments}
We thank Hrvoje Buljan, Robert Pezer, Osor Bari\v{s}i\'{c} and Ivan Balog for very useful suggestions and discussions. This work was supported  by the QuantiXLie Centre of Excellence, a project cofinanced by the Croatian Government and European Union through the European Regional Development
Fund - the Competitiveness and Cohesion Operational Programme (Grant KK.01.1.1.01.0004).
\end{acknowledgments}

\end{document}